\documentstyle[aps,prbbib]{revtex}

\begin{document}

\title{Thermal Magnetization Reversal in Arrays of Nanoparticles}
\author{Gregory Brown}
\address{
School of Computational Science and Information Technology,
Center for Materials Research and Technology,
and Department of Physics,
Florida State University, Tallahassee, Florida 32306-4120
}
\author{M. A. Novotny}
\address{
School of Computational Science and Information Technology,
Florida State University, Tallahassee, Florida 32306-4120
}
\author{Per Arne Rikvold}
\address{
Center for Materials Research and Technology,
School of Computational Science and Information Technology,
and Department of Physics,
Florida State University, Tallahassee, Florida 32306-4350
}

\date{\today}
\maketitle

\begin{abstract}
\centerline{}

The results of large-scale simulations investigating the dynamics of
magnetization reversal in arrays of single-domain nanomagnets after a
rapid reversal of the applied field at nonzero temperature are
presented. The numerical micromagnetic approach uses the
Landau-Lifshitz-Gilbert equation including contributions from thermal
fluctuations and long-range dipole-dipole demagnetizing effects
implemented using a fast-multipole expansion. The individual model
nanomagnets are $9\,{\rm nm} \times 9\,{\rm nm} \times 150\,{\rm nm}$
iron pillars similar to those fabricated on a surface with
STM-assisted chemical vapor deposition [S. Wirth, {\em et al.\/},
J. Appl. Phys {\bf 85}, 5249 (1999)].  Nanomagnets oriented
perpendicular to the surface and spaced $300\,{\rm nm}$ apart in
linear arrays are considered.  The applied field is always oriented
perpendicular to the surface. When the magnitude of the applied field
is less than the coercive value, about $2000\,{\rm Oe}$ for an
individual nanomagnet, magnetization reversal in the nanomagnets can
only occur by thermally activated processes.  Even though the
interaction from the dipole moment of neighboring magnets in this
geometry is only about $1\,{\rm Oe}$, less than $1\%$ of the coercive
field, it can have a large impact on the switching dynamics. What
determines the height of the free-energy barrier is the difference
between the coercive and applied fields, and $1\,{\rm Oe}$ can be a
significant fraction of that. The magnetic orientations of the
neighbors are seen to change the behavior of the nanomagnets in the
array significantly.
\end{abstract}

\vskip 0.15in

The ability of the magnetization to maintain one particular
orientation among many is an essential part of numerous applications
of magnetic materials.  The coercive field is defined to be the
weakest magnetic field for which the magnetization will
deterministically align with the field. For applied magnetic fields
weaker than the coercive field, a free-energy barrier may separate the
orientation of the magnetization from that of the applied magnetic
field. In these weak fields the changes in the orientation of the
magnetization occur only in the unlikely event of thermal crossing of
the free-energy barrier, and magnetization switching becomes a
probabilistic process. In magnetic storage, thermal magnetization
switching is important for understanding the reliability of recorded
data in the presence of stray fields, as well as for thermally
assisted reading and writing. \cite{RUIGROK}

Here we present numerical results for magnetization switching in
linear arrays of weakly coupled nanoscale magnetic pillars, with each
pillar's long axis oriented in the $z$-direction, perpendicular to the
substrate. The simulations start at $t$$=$$-0.25\,{\rm ns}$ with zero
external field and the average $z$-component of the magnetization,
$M_z,$ oriented in the positive $z$-direction. The external field is
then applied according to $H_z(t)=-H_0 \cos{(2\pi t/1\,{\rm ns})}$ for
$-0.25\,{\rm ns} < t < 0\,{\rm ns},$ giving a final value of $-H_0.$

The single-crystal nanomagnet pillars considered here are modeled
after iron nanopillars constructed using STM-assisted chemical vapor
deposition.  \cite{WIRTH98,WIRTH99} The numerical model consists of
magnetization vectors of magnitude unity on a cubic lattice, ${\bf
M}({\bf r}_i)$, with the motion of the vectors given by the
Landau-Lifshitz-Gilbert equation, \cite{BROW63,AHARONI}
\begin{equation}
\label{eq:rllg}
\frac{{d}{\bf M}({\bf r}_i)}{{d}t} =
\frac{\gamma_0}{1+\alpha^2}{\bf M}({\bf r}_i) \times
\left[
  {\bf H}({\bf r}_i) 
  - {\alpha} {\bf M}({\bf r}_i) \times {\bf H}({\bf r}_i)
\right]
\;,
\end{equation}
where $\gamma_0$ is the gyromagnetic ratio $1.76 \times 10^7\, {\rm
Hz/Oe}$ and ${\bf H}({\bf r}_i)$ is the local field at each site. The
local fields have contributions corresponding to exchange,
dipole-dipole interactions, and random thermal noise.  Underdamped
precession of the spins is selected by taking the damping parameter
$\alpha=0.1$; the other material parameters were chosen to match those
of bulk iron, and are $3.6\,{\rm nm}$ for the exchange length,
$1700\,{\rm emu/cm^3}$ for the saturation magnetization, and zero
crystalline anisotropy. Details of the numerical model will appear in
Ref.~[6]. Two numerical models of the nanopillars are considered: a
large-scale simulation with each pillar modeled with $4949$ vectors and
a simpler model with each pillar modeled with only $17$ vectors.

The large-scale simulations model each $9\,{\rm nm} \times 9\,{\rm nm}
\times 150\,{\rm nm}$ pillar using a $7$ $\times$ $7$ $\times$ $101$
lattice.  For these simulations, the time-consuming dipole-dipole
calculations were performed using the fast-multipole method,
\cite{UNPUB,GREE87} and the simulations were run on a series of
massively parallel computers including a CRAY T3E and two different IBM
SP's. The results reported here represent $10^5$ cpu-hours of
computation.  The pillars were separated by twice their length, or
$300\,{\rm nm}.$ At this separation the interactions between
neighboring iron pillars are on the order of $1\,{\rm Oe}.$

Here we consider linear arrays of nanopillars, $4 \times 1$ arrays in
which there are two classes of pillars, which we call ``inside'' and
``outside.''  The switching time for each pillar is taken to be the
first time the value of $M_z$ for that pillar passes through zero, and
it is measured from $t$$=$$0$. The results are presented in Fig.~1 as
$P_{\rm not}(t),$ the probability that a pillar has not switched up to
time $t.$ A total of $30$ array switches, thus $60$ for each class of
pillar, were simulated for $H_0$$=$$1800\,{\rm Oe}$ and
$T$$=$$20\,{\rm K},$ and the results are compared to $100$ switches
simulated for isolated pillars under the same conditions.
\cite{UNPUB,JAP} The field magnitude used here is below the coercive
field for isolated nanoparticles, which has been estimated to be $1995
\pm 20 \,{\rm Oe}$ in dynamic, nonequilibrium simulations where the
field is swept linearly. \cite{UNPUB} From these results it appears
that the outside pillars switch, on average, at earlier times than
either the inside pillars of the array or the isolated pillars. We
note that functional forms for $P_{\rm not}(t)$ that are neither
exponentials nor the error functions associated with Gaussian
statistics have been observed experimentally.  \cite{LEDERMAN,KOCH} A
simple theory with an analytic form for $P_{\rm not}$ reasonably
describes the simulation results for isolated pillars.
\cite{UNPUB,JAP}

To investigate switching involving much larger free-energy barriers,
and thus longer switching times, simulations were also conducted with
a simplified model of the pillars, specifically with each pillar
modeled by a $1$ $\times$ $1$ $\times$ $17$ arrangement of
magnetization vectors. \cite{BandB} This models a $5.2\,{\rm nm}
\times 5.2\,{\rm nm} \times 88.4\,{\rm nm}$ iron pillar, which has
approximately the same aspect ratio as the previous pillar. For an
isolated pillar of this type, the coercive field is approximately
$1500\,{\rm Oe}.$ The pillars in these arrays were separated by
$176.8\,{\rm nm},$ and the dipole-dipole interactions were calculated
using direct summation. \cite{UNPUB,BandB} Because this approach is
$O(N^2)$ the calculation, which is fast for individual pillars, is
slow for even small arrays. Still, these simulations are
computationally less expensive than those discussed above, and results
are presented for $1269$ array switches and $1986$ isolated pillar
switches. The results reported here required about $10^4$ cpu-hours of
computation on desktop workstations.

The $P_{\rm not}(t)$ for $4 \times 1$ arrays of the simple model with
$H_0=1000\,{\rm Oe}$ and $T=20\,{\rm K}$ are shown in Fig.~2, along
with the results for isolated pillars under the same conditions. In
this case, the differences in $P_{\rm not}(t)$ are small, but the
array pillars still appear to have a small bias to switch at earlier
times than isolated pillars. Under these conditions, however, the
effect of the magnetic orientation of the nearest-neighbor pillars can
be seen.  Fig.~3 shows subsets of $P_{\rm not}(t)$ for the switching
of inside pillars in the simple model. The dotted and dashed curves
are $P_{\rm not}$ for inside pillars where one and both
nearest-neighbor pillars have already switched, respectively. Here the
switching time is measured from the last time one of the
nearest-neighbor pillars switched. The data is presented on a
linear-log scale, and in both cases $P_{\rm not}$ is consistent with
an exponential form. This suggests that pillars in these environments
have an approximately constant, history-independent decay rate
determined by the orientation of the nearest-neighbor pillars. The
exponential form is consistent with our simple theory for $P_{\rm
not}$ in nanomagnet pillars, \cite{UNPUB,JAP} because environments
with switched neighbors cannot exist for times less than $t_{\rm g},$
the growth time from nucleation of one endcap to $M_z$ passing through
zero. The solid curve is $P_{\rm not}$ constructed from the data for
pillars with two neighbors, neither of which have switched.  The
$t_{\rm sw}$ used for this data occur at times greater than $t_g,$
here estimated using our simple theory \cite{UNPUB,JAP} as twice the
earliest observed $t_{\rm sw}.$ In addition $t_g$ has been subtracted
from each $t_{\rm sw}$ for this curve. Similar results for the outside
pillars of the array are shown in the inset of Fig.~3. The trend
towards slower decay with increasing number of switched neighbors is
consistent with a simple picture of dipole-dipole interaction between
nearest-neighbor pillars with an unswitched neighbor contributing a
field parallel to the external field and a switched neighbor
contributing a field antiparallel to the external field. Currently, a
sufficient number of samples do not exist to reliably repeat this
analysis for the larger simulations.

To summarize, we have simulated the switching dynamics in linear
arrays of nanomagnet pillars after a reorientation of the external
field, both for a simple model in which each pillar is represented by
a one-dimensional array of magnetization vectors and for a large-scale
model with each pillar represented by a three-dimensional lattice of
vectors. In one-dimensional arrays of pillars that are mutually
located in the other pillars' far fields and that have small
free-energy barriers, the $P_{\rm not}(t)$ of the pillars at the end
of the arrays falls off faster than that of isolated pillars and the
inside pillars of the array. For pillars that have large free-energy
barriers, the long switching-time tails of $P_{\rm not}$ for pillars
with different numbers of switched nearest neighbors show exponential
behavior, with decay constants that decrease with increasing numbers
of switched neighbors.  This occurs because the presence of unswitched
(switched) neighbors enhances (retards) switching.

Supported by NSF grant No. DMR-9871455, NERSC, and by FSU/CSIT,
FSU/MARTECH, and FSU/ACNS.

~
\begin{figure}
\null
\vskip 2.00in
\includegraphics{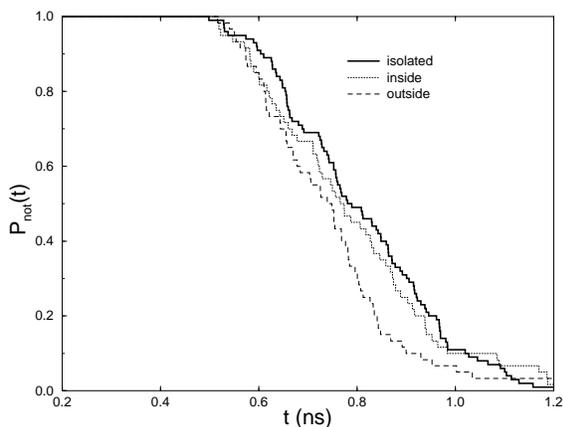}
\caption[]
{Probability of not switching, $P_{\rm not}$, for
large-scale simulations of nanomagnet pillars in a $4 \times 1$
array. The solid curve is for isolated pillars, the dotted curve
is for inside pillars, and the dashed curve is for outside pillars for
30 simulated array switches and $100$ simulated isolated-pillar
switches with $H_0$$=$$1800$~Oe and $T$$=$$20$~K.}
\end{figure}

~
\begin{figure}
\null
\vskip 2.00in
\includegraphics{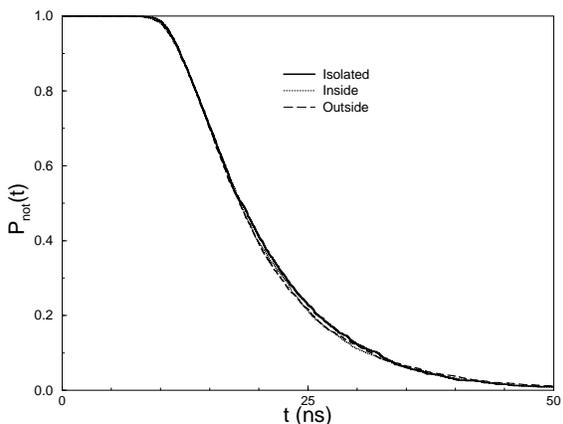}
\caption[]
{Probability of not switching, $P_{\rm not}$,
for simulations of a simple model of nanomagnets in a $4 \times 1$
array. The solid curve is for isolated pillars, the dotted curve
is for inside pillars, and the dashed curve is for outside
pillars. These results are for $1269$ simulated array switches and
$1986$ simulated isolated-pillar switches at $H_0$$=$$1000$~Oe and
$T$$=$$20$~K. The free-energy barrier in this reversal is higher than
for the results shown in Fig.~1.}
\end{figure}

\newpage
~
\begin{figure}
\null
\vskip 2.00in
\includegraphics{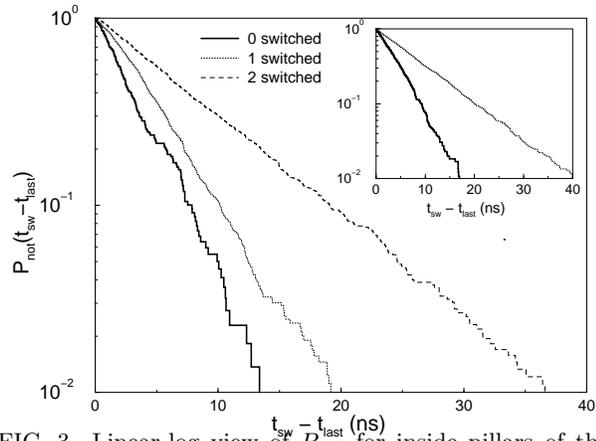}
\caption[]
{Linear-log view of $P_{\rm not}$ for inside pillars of the
simple model with different orientations of the nearest-neighbor
pillars.  The dotted and dashed curves are for one and both neighbors
switched, respectively, and the switching time, $t_{\rm sw},$ is
measured since the last time a neighbor switched. The solid curve is
the exponential tail of $P_{\rm not}$ for pillars with both neighbors
unswitched, shifted along the $x$-axis by the growth time, $t_g.$
Similar results are shown for the outside pillars in the inset.  }
\end{figure}

\end{document}